\documentclass[aps,prd,eqsecnum,nofootinbib,showpacs,twocolumn]{revtex4-2}
\usepackage{amsmath,amssymb,bm}
\usepackage[dvipdfmx]{graphicx}
\usepackage{color}

\numberwithin{equation}{section}

\newcommand{\be}{\begin{equation}}
\newcommand{\ee}{\end{equation}}
\newcommand{\bea}{\begin{eqnarray}}
\newcommand{\eea}{\end{eqnarray}}

\newcommand{\eq}[1]{Eq.~(\ref{eq:#1})}
\newcommand{\sect}[1]{Sec.~\ref{sec:#1}}
\newcommand{\appen}[1]{Appendix~\ref{sec:#1}}

\newcommand{\del}{\partial}

\newcommand{\eg}{{\it e.g.}}
\newcommand{\ie}{{\it i.e.}}

\newcommand{\bh}{black hole\ }

\bmdefine{\bmq}{{\bm{q}}}
\bmdefine{\bmk}{{\bm{k}}}
\bmdefine{\bmx}{{\bm{x}}}
\bmdefine{\bmy}{{\bm{y}}}
\bmdefine{\bmr}{{\bm{r}}}
\bmdefine{\bmnabla}{{\bm{\nabla}}}
\bmdefine{\bmA}{ \bm{A} }
\bmdefine{\bmD}{ \bm{D} }

\bmdefine{\bmPhi}{ \bm{\Phi} }
\bmdefine{\bmPsi}{ \bm{\Psi} }
\bmdefine{\bmcalO}{ \bm{\mathcal{O}} }

\newcommand{\calM}{{\cal M}}

\newcommand{\nq}{\mathfrak{q}}

\newcommand{\nw}{\mathfrak{w}}

\newcommand{\calP}{{\cal P}}

\bmdefine{\bmg}{{\bm{g}}}
\bmdefine{\bmR}{{\bm{R}}}

\newcommand{\mfh}{\mathfrak{h}}

\newcommand{\mfA}{\mathfrak{A}}

\newcommand{\twonabla}{{}^{(2)}\! \nabla}

\newcommand{\half}{\frac{1}{2}}
\newcommand{\schrodinger}{Schr\"{o}dinger}

\newcommand{\bwt}{\begin{widetext}}
\newcommand{\ewt}{\end{widetext}}
\newcommand{\bab}{\begin{autobreak}}
\newcommand{\eab}{\end{autobreak}}

\newcommand{\nk}{\mathfrak{p}}
\newcommand{\ki}{p}
\newcommand{\tilZ}{\tilde{Z}}
\newcommand{\vecX}{\vec{X}}
\newcommand{\veca}{\vec{a}}

\newcommand{\hatt}{\hat{t}}
\newcommand{\hz}{\hat{z}}
\newcommand{\homega}{\hat{\omega}}
\newcommand{\hq}{\hat{q}}
\newcommand{\hnq}{\hat{\nq}}
\newcommand{\Za}{\tilZ}

\newcommand{\vphi}{\varphi}

\newcommand{\boxeq}[1]{
{\fboxsep=10pt \fbox{$\displaystyle #1 $}}
}

\begin{document}

%\西暦
%\begin{flushright}
%        \today\\
%        {\it Preliminary version}
%\end{flushright}

%%%   Title Page   %%%

\title{Pole-skipping in a non-black-hole geometry}
\author{Makoto Natsuume}
\email{makoto.natsuume@kek.jp}
\altaffiliation[Also at]{
Graduate Institute for Advanced Studies, SOKENDAI,
1-1 Oho, Tsukuba, Ibaraki, 305-0801, Japan;
 Department of Physics Engineering, Mie University, 
 Tsu, 514-8507, Japan.}
\affiliation{KEK Theory Center, Institute of Particle and Nuclear Studies, 
High Energy Accelerator Research Organization,
Tsukuba, Ibaraki, 305-0801, Japan}
\author{Takashi Okamura}
\email{tokamura@kwansei.ac.jp}
\affiliation{Department of Physics and Astronomy, Kwansei Gakuin University,
Sanda, Hyogo, 669-1330, Japan}
\date{\today}
\begin{abstract}
The pole-skipping has been discussed in black hole backgrounds, but we point out that the pole-skipping exists even in a non-black-hole background, the AdS soliton. For black holes, the pole-skipping points are typically located at imaginary Matsubara frequencies $\omega=-(2\pi T)ni$ with an integer $n$. The AdS soliton is obtained by the double Wick rotation from a black hole. As a result, the pole-skipping points are located at $q_z=-(2\pi n)/l$, where $l$ is the $S^1$ periodicity and $q_z$ is the $S^1$ momentum. The ``chaotic" and the ``hydrodynamic" pole-skipping points lie in the physical region. We also propose a method to identify all pole-skipping points instead of the conventional method.
\end{abstract}
%The pole-skipping phenomena have been discussed in black hole backgrounds, but we study the pole-skipping in a non-black-hole background, the AdS soliton. 
%
%\pacs{11.25.Tq, 64.60Ht, 25.75.-q}%, 74.20.-z} %?
% KEK-TH-2521

\maketitle

%%%%%%%%%
\section{Introduction and Summary}%\label{sec:}
%%%%%%%%%

The retarded Green's functions play fundamental roles in physics, and considerable knowledge 
% a great deal of, extensive
has been accumulated over the years. However, surprisingly, a new universal feature of the Green's functions are found using the AdS/CFT duality or holographic duality \cite{Maldacena:1997re,Witten:1998qj,Witten:1998zw,Gubser:1998bc} (see, \eg, Refs.~\cite{CasalderreySolana:2011us,Natsuume:2014sfa,Ammon:2015wua,Zaanen:2015oix,Hartnoll:2016apf,Baggioli:2019rrs}). 
This feature is known as the \textit{pole-skipping} \cite{Grozdanov:2017ajz,Blake:2018leo,Grozdanov:2019uhi,Blake:2019otz,Natsuume:2019xcy}. 
% v1
%This feature is known as the pole-skipping \cite{Grozdanov:2019uhi,Blake:2019otz,Natsuume:2019xcy}. 
% v2 alternative
%This feature is known as the \textit{pole-skipping}. The pole-skipping was first found in the context of holographic chaos \cite{Grozdanov:2017ajz,Blake:2018leo}, but it was soon realized that various Green's functions show the universal behavior \cite{Grozdanov:2019uhi,Blake:2019otz,Natsuume:2019xcy}.
Since then, various aspects of the pole-skipping have been investigated (see, \eg, Refs.~\cite{Natsuume:2019sfp,Natsuume:2019vcv,Wu:2019esr,Balm:2019dxk,Ceplak:2019ymw,Liu:2020yaf,Ahn:2019rnq,Ahn:2020bks,Abbasi:2020ykq,Jansen:2020hfd,Ramirez:2020qer,Ahn:2020baf,Natsuume:2020snz,Kim:2020url,Sil:2020jhr,Ceplak:2021efc,Jeong:2021zhz,Natsuume:2021fhn,Blake:2021hjj,Kim:2021xdz,Wang:2022mcq,Amano:2022mlu,Yuan:2023tft,Grozdanov:2023txs}).

Typical bulk perturbation problems are the scalar field, the Maxwell field, and the gravitational field. The dual Green's functions are not uniquely determined at ``pole-skipping points" in the complex momentum space $(\omega,q)$ where $\omega$ is frequency and $q$ is wave number. Near a pole-skipping point, the Green's function typically takes the form
\begin{align}
G^R \propto \frac{\delta\omega+\delta q}{\delta\omega-\delta q}~.
%\label{eq:}
%
\end{align}
In this sense, the Green's function is not uniquely determined, and it depends on the slope $\delta q/\delta\omega$ how one approaches the pole-skipping point. The hydrodynamic pole is an elementary example. For example, the sound mode behaves as
\begin{align}
G^R \propto \frac{q^2}{3\omega^2-q^2} \to \frac{\delta (q^2)}{3\delta(\omega^2)-\delta(q^2)}~.
%\label{eq:}
%
\end{align}
From the bulk point of view, the pole-skipping occurs because the bulk solution is not uniquely determined there. 

There is a universality for the pole-skipping points $\omega$. 
In these examples, the pole-skipping points are located at Matsubara frequencies.%
\footnote{A rotating black hole is an exception \cite{Natsuume:2020snz}. For the rotating BTZ black hole, the pole-skipping points $\omega$ depend on the left-moving temperature $T_L$, the right-moving temperature $T_R$, and the conformal dimension $\Delta_+$ of the dual boundary operator.}
The pole-skipping points start from $\nw := \omega/(2\pi T) =(s-1)i $ ($s$ is the spin of the bulk field) 
and continue to $\nw_n = (s-1-n)i $ for a non-negative integer $n$. Namely, for the scalar field, they start from $\nw_1=-i$. For the Maxwell field, they start from $\nw_0=0$ which is the hydrodynamic pole. In the gravitational sound mode, they start from $\nw_{-1}=+i$. It is argued that the $\nw_{-1} = +i$ point is related to many-body quantum chaos \cite{Shenker:2013pqa,Roberts:2014isa,Roberts:2014ifa,Shenker:2014cwa,Maldacena:2015waa}.

The pole-skipping was first found in the context of holographic chaos \cite{Grozdanov:2017ajz,Blake:2018leo}, so one naturally has studied the pole-skipping at finite temperature or in a \bh background. However, an analogous phenomenon was found even in elementary quantum mechanics problem \cite{Natsuume:2021fhn}, so one may discuss the phenomena in a broader context. 

In this paper, we study the pole-skipping in a non-black-hole geometry, the AdS soliton \cite{Horowitz:1998ha}. The AdS soliton is obtained by the double Wick rotation from the AdS black hole (\sect{soliton}). The AdS soliton is not a black hole. The geometry has a compact $S^1$-direction $z$ with periodicity $l$, and the geometry ends smoothly at the ``horizon." The AdS \bh describes the plasma phase of a large-$N_c$ gauge theory, whereas the AdS soliton describes the confining phase.
%For the SAdS$_5$ black hole, the imaginary time direction has the periodicity $\beta=\pi/r_0$ to avoid a conical singularity. Similarly, for the AdS soliton, $z'$ has the periodicity $l=\pi/r_0$. The AdS soliton is not a black hole. Rather, the geometry ends smoothly at $r=r_0$. The AdS black hole describes the plasma phase whereas the AdS soliton describes the confining phase. 

Our results are summarized as follows:
\begin{enumerate}
\item
Because of the double Wick rotation, the universality of the pole-skipping points $\omega$ is translated to the universality of the pole-skipping points $\nq_z:=q_z/(2\pi/l)$, where $q_z$ is the wave number in the $S^1$-direction $z$.  
\item
The pole-skipping points start from $\nq_z=(s-1)$ and continue to $\nq_z=-n$. 
While some pole-skipping points lie in the physical region (\sect{pole_skip_soliton}), most pole-skipping points do not lie in the physical region. The former corresponds to the pole-skipping points in the upper-half $\omega$-plane in the \bh case, namely the ``chaotic" mode and the ``hydrodynamic" mode, 
%The pole-skipping points which correspond to the lower-half $\omega$-plane for a \bh do not lie in the physical region (\sect{pole_skip_soliton}), but  the pole-skipping points which correspond to the upper-half $\omega$-plane, namely the ``chaotic" mode and the ``hydrodynamic" mode, lie in the physical region. 
Of course, the AdS soliton is not a black hole, so the ``chaotic" mode does not imply a chaotic behavior.
\item 
Because $q_z$ is the $S^1$-momentum, there is also the mirror image of the pole-skipping tower. Namely,  they start from $\nq_z=-(s-1)$ and continue to $\nq_z=n$.
\end{enumerate}
We discuss the bulk scalar field, the Maxwell field, and the gravitational field. There is a conventional pole-skipping method, but we propose an alternative method in order to analyze the pole-skipping systematically (\sect{pole_skip}). The conventional method requires a separate treatment for the chaotic and hydrodynamic pole-skippings, but our method covers them as well. The perturbation problems in the AdS soliton with $\nq_z\neq0$ are little discussed in the literature, and our work is interesting from this point of view as well.

%%%%%%%%%
\section{Pole-skipping}\label{sec:pole_skip}
%%%%%%%%%

In this section, we briefly review the pole-skipping in the context of the Schwarzschild-AdS$_5$ (SAdS$_5$) black hole:
\begin{subequations}
\label{eq:sads5}
\begin{align}
ds_5^2 &= r^2(-fdt^2+dx^2+dy^2+dz^2)+\frac{dr^2}{r^2f} \\
&= \frac{r_0^2}{u} (-fdt^2+dx^2+dy^2+dz^2)+\frac{du^2}{4u^2f}~, \\
f &= 1-\left(\frac{r_0}{r}\right)^4 = 1-u^2~, 
%\label{eq:}
%
\end{align}
\end{subequations}
where $u:=r_0^2/r^2$.
For simplicity, we set the AdS radius $L=1$ and the horizon radius $r_0=1$. The Hawking temperature is given by $\pi T =r_0/L^2$. 

We consider the perturbation of the form
\begin{align}
Z(u) e^{-i\omega t+iqx}~.
%\label{eq:}
%
\end{align}
Near the horizon, the field equation typically takes the form
\begin{align}
0 \sim Z''+\frac{1}{u-1}Z'+\frac{\nw^2}{4(u-1)^2}Z~, \quad(u\to1)~,
%\label{eq:}
%
\end{align}
where $'=\del_u$ and $\nw:=\omega/(2\pi T)=\omega/2$, so the solution behaves as
\begin{align}
Z \propto (u-1)^{\pm i\nw/2}~.
%\label{eq:}
%
\end{align}
As usual, we impose the ``incoming-wave" boundary condition at the horizon, so we set the ansatz
\begin{align}
Z = (u-1)^{-i\nw/2}\Za~.
%\label{eq:}
%
\end{align}

As a typical example of pole-skipping, we consider the field equation of the form
\begin{align}
0= \Za''+P(u)\Za'+Q(u)\Za~.
%\label{eq:}
%
\end{align}
The horizon $u=1$ is a regular singularity, and $P$ and $Q$ are expanded as
\begin{subequations}
%\label{eq:}
\begin{align}
P &= \frac{P_{-1}}{u-1} + P_0 + P_1(u-1) + \cdots~, \\
Q &= \frac{Q_{-1}}{u-1} + Q_0 + Q_1(u-1) + \cdots~.
%\label{eq:}
%
\end{align}
\end{subequations}
One typically has $P_{-1}=1-i\nw$ and $Q_{-1}=Q_{-1}(\nw,\nq^2)$, where $\nq:=q/(2\pi T)$.

We first review a conventional method \cite{Blake:2019otz} and then propose an alternative method. 

%%------------------
\subsection{Conventional method}%\label{sec:}
%%------------------

The solution can be written as a power series:
\begin{align}
\Za(u) = \sum_{n=0}\, a_n\, (u-1)^{n+\lambda}~.
%\label{eq:}
%
\end{align}
Substituting this into the field equation, one obtains the indicial equation at the lowest order: 
\begin{align}
\lambda=0~, \quad 1-P_{-1}~.
%\label{eq:}
%
\end{align}
The coefficient $a_n$ is obtained by a recursion relation. The mode with $\lambda=0$ ($\lambda=1-P_{-1}=i\nw$) represents the incoming (outgoing) mode, and we choose the incoming mode $\lambda=0$ hereafter. 

In order to obtain pole-skipping points systematically, write the rest of the field equation in a matrix form \cite{Blake:2019otz}:
\begin{subequations}
%\label{eq:}
\begin{align}
0 & = M\Za \\
& =\begin{pmatrix} 
    M_{11} & M_{12} & 0 & 0 & \cdots \\
    M_{21} & M_{22} & M_{23} & 0 & \cdots \\
    \cdots & \cdots & \cdots & \cdots & \cdots 
  \end{pmatrix}
  \begin{pmatrix} 
    a_0 \\ a_1 \\ \cdots
  \end{pmatrix}~.
\label{eq:recursion_matrix}
\end{align}
\end{subequations}

One can show that $M_{n,n+1} = n(n-1+P_{-1})=n(n-i\nw)$. The matrix $\calM^{(n)}$ is obtained by keeping the first $n$ rows and $n$ columns of $M$. The pole-skipping points are obtained by
\begin{align}
\boxeq{
M_{n,n+1}(\nw_n) = 0~, \quad
\det \calM^{(n)}(\nw_n,\nq_n) = 0~.
}
\label{eq:condition1}
\end{align}

%\be
%
%\fbox{$
%\begin{aligned}
%M_{n,n+1}(\nw_n) &= 0  \\
%\det \calM^{(n)}(\nw_n,\nq_n) &= 0
%\end{aligned}
%$}
%\label{eq:bulk_bdy_mapping}
%
%\ee

For example, consider the first row:
\begin{align}
0 = M_{11} a_0 + M_{12}a_1 = Q_{-1}a_0+P_{-1}a_1 = 0~.
%\label{eq:}
%
\end{align}
Normally, this equation determines $a_1$ from $a_0$. However, when $M_{12}=M_{11}=0$ or $P_{-1}=Q_{-1}=0$, both $a_0$ and $a_1$ are free parameters, and the bulk solution is not uniquely determined. 

Similarly, when $M_{23}=\det \calM^{(2)}=0$, $a_0$ and $a_2$ become free parameters. One gets 
\begin{subequations}
\label{eq:pole_skip_2nd}
\begin{align}
M_{23} &= 2(2-i\nw)=0~, \\
\det \calM^{(2)} &= Q_{-1}(Q_{-1}+P_0)-P_{-1}Q_0=0~.
\end{align}
\end{subequations}

As is clear from the construction, one can find pole-skipping points only in the lower-half $\omega$-plane $\nw_n=-in (n>0)$ in this method. The ``chaotic" pole-skipping $n=-1$ and the ``hydrodynamic" pole-skipping $n=0$ 
% v2.1
need a separate treatment as pointed out in Ref.~\cite{Blake:2019otz}. In the next subsection, we propose an alternative method where one can find all pole-skipping points $\nw_n=-in (n\geq-1)$.

%%------------------
\subsection{Alternative method}\label{sec:matrix_formalism}
%%------------------

%%------------------
\subsubsection{The formalism}%\label{sec:}
%%------------------

We start to write the field equation in a matrix form:
\begin{subequations}
%\label{eq:}
\begin{align}
0 &= \vecX'- M \vecX~, \\
\vecX &:= \begin{pmatrix} 
\Za \\ 
\Za'
 \end{pmatrix}~,
\\
M &:= 
  \begin{pmatrix} 
  0 &1 \\
  -Q & -P
  \end{pmatrix}~.
%\label{eq:}
%
\end{align}
\end{subequations}
The matrix $M$ can be expanded as
\begin{align}
M= \frac{M_{-1}}{u-1}+M_0+M_1(u-1)\cdots~.
%\label{eq:}
%
\end{align}
The solution can be written as a power series:
\begin{align}
\vecX = \sum_{n=0}\, \veca_n\, (u-1)^{n+\lambda}~.
%\label{eq:}
%
\end{align}
Substituting this into the field equation, at the lowest order, one obtains
\begin{align}
0 =(\lambda - M_{-1})\veca_0~.
%(\lambda+1 - \calM)\veca_1=m_0\veca_0~.
%\label{eq:}
%
\end{align}
This indicial equation is an eigenvalue equation for $M_{-1}$. 
The eigenvalue and the eigenvector of $M_{-1}$ are 
\begin{subequations}
%\label{eq:}
\begin{align}
  & \lambda = 0~,
& & \veca_0
  = \begin{pmatrix} 
  1 \\
  b_1
    \end{pmatrix}
  ~, \quad b_1= -\frac{Q_{-1}}{P_{-1}}~,
\\
  & \lambda = i\nw -1~,
& & \veca_0
  = \begin{pmatrix} 
  0 \\ 
  1 
  \end{pmatrix}
  ~.
\end{align}
\end{subequations}
The mode with $\lambda=0$ ($\lambda=i\nw-1$) represents the incoming (outgoing) mode, and we choose the incoming mode $\lambda=0$ hereafter. 

 In this matrix formalism, one looks for the points where the coefficient $\veca_n$ becomes ambiguous. 
 The functions $P_{-1}$ and $Q_{-1}$ are polynomials in $(\nw,\nq^2)$, so $b_1$ is a rational function of $(\nw,\nq^2)$. Then, $b_1$ becomes ambiguous or becomes $0/0$ when $P_{-1}=Q_{-1}=0$. Then, the bulk solution is not uniquely determined because the coefficient of the Frobenius series becomes ambiguous. This agrees with the first pole-skipping point in the conventional formalism in the last subsection.

Once one obtains $\veca_0$, $\veca_n$ is obtained recursively:
\begin{align}
(\lambda+n-M_{-1})\veca_n = \sum_{k=0}^{n-1} M_{n-1-k}\veca_k~,
%\label{eq:}
%
\end{align}
with $\lambda=0$. 

For example, at the next order,
\begin{subequations}
%\label{eq:}
\begin{align}
(1-M_{-1})\veca_1 &= M_0 \veca_0~, \\
\veca_1
  &= \begin{pmatrix} 
  b_1 \\
  2b_2
    \end{pmatrix}
~, \\
2b_2 &= \frac{ Q_{-1}(Q_{-1}+P_0)-P_{-1}Q_0 }{ P_{-1}(P_{-1}+1) }~.
%\label{eq:}
%
\end{align}
\end{subequations}
Thus, $b_2$ becomes ambiguous when
\begin{align}
P_{-1}+1=0~, \quad
Q_{-1}(Q_{-1}+P_0)-P_{-1}Q_0=0~.
%\label{eq:}
%
\end{align}
This agrees with the second pole-skipping point in the conventional method \eqref{eq:pole_skip_2nd}. 

In general, the coefficient $\veca_n$ can be written as
\begin{align}
\veca_n
  = \begin{pmatrix} 
  b_n\\
  (n+1)b_{n+1}
    \end{pmatrix}
%\label{eq:}
%
\end{align}
with $b_0=1$. The pole-skipping condition is
\begin{align}
\boxeq{
b_n(\nw_{n},\nq_{n})=\frac{0}{0}~.
}\label{eq:condition2}
\end{align}

%%------------------
\subsubsection{Advantages}%\label{sec:}
%%------------------

The matrix formalism has 2 advantages compared with the conventional method. 

First, as discussed above, this formalism gives the equivalent results to the conventional formalism, but there are important exceptions. When $Q_{-1}=0/0$, $b_1$ becomes ambiguous as well. This is the case for the ``chaotic" pole-skipping and the ``hydrodynamic" pole-skipping. We show this explicitly in the context of the AdS soliton (\sect{pole_skip_soliton}). Of course, to obtain all pole-skipping points, one would use the conventional formalism and impose the condition $Q_{-1}=0/0$ separately. But in this matrix formalism, all pole-skipping points are included naturally. 

Namely, there are 2 kinds of pole-skipping:
\begin{enumerate}
\item
The first one is the pole-skipping in the lower-half $\omega$-plane. They come from the fact that field equations have a regular singularity at the horizon. In the conventional method, two roots of the indicial equation are $\lambda=0,i\nw$. $\lambda$ depends on $\nw$ because of a regular singularity, and the pole-skipping occurs when they differ by an integer.
% If field equations have only ordinary points, this does not occur: two roots are $\lambda=0,1$ and their difference is not $\omega$-dependent.
\item
The second one is the pole-skipping in the upper-half $\omega$-plane, namely the chaotic and hydrodynamic pole-skipping. In this case, the coefficients of the field equations themselves partly have the $0/0$ structure. 

\end{enumerate}

Second, the conventional method is applicable if one can find a master equation, but the matrix formalism is applicable even if one cannot find a master equation:
\begin{enumerate}
\item
It is often not easy to find a master field.
\item
Even if one finds a master field, the choice of a master field is not unique. This is particularly problematic for the pole-skipping analysis. As discussed in \appen{different}, one is not able to find some pole-skipping points if one chooses a different master variable.
\end{enumerate}
The matrix formalism does not have such disadvantages. In perturbation problems, one typically has coupled first-order differential equations with a multiple number of variables. For example, the Maxwell scalar mode (\sect{diffusive}) consists of 2 coupled first-order differential equations with 2 variables $(\mfA_z,\mfA_u)$. In this case, it is easy to find the master equation for the master field $\mfA_z$. But if one chooses the master field $\mfA_u$, one is not able to find the ``hydrodynamic" pole-skipping (\appen{different}).

In this case, it is not really necessary to use a master equation. It is enough to rewrite the coupled equations in a matrix form setting $\vecX$ as
\begin{align}
\vecX &:= \begin{pmatrix} 
\mfA_z \\ 
\mfA_u
 \end{pmatrix}~.
%\label{eq:}
%
\end{align}

%%------------------
\subsubsection{A subtlety}%\label{sec:}
%%------------------

% v2
There is a subtlety to use the matrix formalism for higher $n$. For example, consider $b_2$. In general, $b_2$ takes a complicated form, and its numerator may have an accidental zero at the first pole-skipping point $n=1$. Such points are not pole-skipping points and should be excluded. In fact, one can easily check that the eigenvector $\veca_1$ does not have the slope-dependence. $\veca_1$ is given by
\begin{align}
\veca_1
  = \begin{pmatrix} 
  b_1\\
  2b_2
    \end{pmatrix}~,
%\label{eq:}
%
\end{align}
and $b_1$ determines the first pole-skipping point. The accidental zero is not included in $b_1$, so $b_1$ diverges there.
%If one normalize $\veca_0$ appropriately, this is the ``outgoing mode" we throw away.  Also, one can show that the additional zero gives the pole-skipping point for the outgoing mode. 
Also, such accidental ones in general do not appear at higher orders.  We show this explicitly in the context of the AdS soliton (\sect{pole_skip_soliton}). To avoid them, a simple rule is
\begin{center}
Find $b_n=0/0$ at $\nw_n$~,
\end{center}
and one should not consider smaller $n$ because the fake ones appear.

We end this section with a list of first pole-skipping points:
\begin{enumerate}
\item Massive scalar: $(\nw,\nq^2) = (-i,-(6+m^2)/4)$.
\item Maxwell vector: $(\nw,\nq^2) = (-i,-1/2)$.
\item Maxwell scalar: $(\nw,\nq^2) = (0,0)~, (-i,+1/2)$.
\item Gravitational vector (shear): \\ 
$(\nw,\nq^2) = (0,0)~, (-i, 3/2)$.
\item Gravitational scalar (sound): \\
$(\nw,\nq^2) =(+i, -3/2)~, (0,0)~,(-i, (1\pm2\sqrt{2}i)/2)$.
\end{enumerate}
The above results can be obtained from the master equations by Kovtun and Starinets \cite{Kovtun:2005ev} (\appen{KS}).

%%%%%%%%%
\section{AdS soliton}\label{sec:soliton}
%%%%%%%%%

The SAdS$_5$ black hole is given by \eq{sads5}. We now compactify the $z$-direction as $0\leq z<l$. However, the compactified solution is not the only geometry with asymptotic geometry $\mathbb{R}^{1,2}\times S^1$. The AdS soliton is also the solution with the same asymptotic geometry. The AdS soliton is obtained by the double Wick rotation from the SAdS$_5$ black hole:
\begin{align}
t=i\hz~, \quad \hatt=iz~.
%\label{eq:}
%
\end{align}
Then, the metric becomes
\begin{align}
ds_5^2 &= \frac{r_0^2}{u} (-d\hatt^2+dx^2+dy^2+fd\hz^2)+\frac{du^2}{4u^2f}~, 
\label{eq:ads_soliton}
\end{align}
with $f=1-u^2$. We set $r_0=1$.

For the SAdS$_5$ black hole, the imaginary time direction has the periodicity $\beta=\pi/r_0$ to avoid a conical singularity. Similarly, for the AdS soliton, $\hz$ has the periodicity $l=\pi/r_0$. The AdS soliton is not a black hole. Rather, it has a cigar-like geometry, and the geometry ends smoothly at $u=1$ because of the factor $f$ just like the Euclidean black hole. We focus on the asymptotically AdS$_5$ geometry, but the generalization to the other dimensions is straightforward.

The SAdS$_5$ black hole describes the plasma phase, whereas the AdS soliton describes the confining phase. Some evidences are
\begin{enumerate}
\item The spectrum has a mass gap \cite{Witten:1998zw,Csaki:1998qr,deMelloKoch:1998vqw}.
\item The quark-antiquark potential is linear and describes the confining potential.
\item The SAdS$_5$ black hole has entropy density $s\sim O(N_c^2)$ whereas the AdS soliton has $s\sim 0$ at leading order in $N_c$ because it is not a black hole.
\end{enumerate}
See \appen{properties} for more details. 
%Below we use $t,z$ instead of $t', z'$.

For the SAdS$_5$, there is a $SO(3)$ invariance for the boundary direction $(x,y,z)$, so one can set the perturbation of the form 
\begin{align}
%
%\label{eq:}
\phi(u) e^{-i\omega t+iqx}
\end{align}
without loss of generality. For the AdS soliton, the invariance is broken due to the $S^1$-direction $z$, so we consider the perturbation of the form
\begin{align}
\phi(u) e^{-i\homega \hatt+iq_x x+i\hq_z \hz}~.
%\label{eq:}
%
\end{align}
However, there is a remaining $SO(1,2)$ invariance for $(\hatt,x,y)$, so $\homega$ and $q_x$ appear only in the combination $\ki^2:=-\homega^2+q_x^2$. 

Because the $z$-direction is compact, $\hq_z$ physically takes only a discrete value:
% v2
\begin{align}
\hq_z = \frac{2\pi n}{l} \text{~~or~~} \nq_z:=\frac{\hq_z}{\frac{2\pi}{l}} =n
%\label{eq:}
%
\end{align}
for an integer $n$. 
But we make it continuous. Such a treatment is often done in the pole-skipping literature. For example, the BTZ black hole has a compact $S^1$-direction $x$, but one makes it noncompact for the pole-skipping analysis. It is similar in spirit to the $S$-matrix approach: even though the angular momentum $l$ is discrete physically, one analytically continues to the whole complex $l$-plane.

%%------------------
\subsection{Expected results}%\label{sec:}
%%------------------

The double Wick rotation $t=i\hz, \hatt=iz$ means that
\begin{align}
i\hq_z = \omega~, iq_z=\homega~.
%\label{eq:}
%
\end{align}
Namely, the role of $\omega$ and $q_z$ is exchanged. Thus, one expects that
\begin{enumerate}
\item
There is no universality for $\homega$ at pole-skipping points. Rather, there is a universality for $\hq_z$.
\item
The pole-skipping points start from $\hnq_z=s-1$, where 
% v2
%$\hnq_z:=\hq_z/(2\pi/l)=\hq_z/2$, and 
$s$ is the spin of the bulk field. The pole-skipping tower continues to $\hnq_z=s-1-n$ where $n$ is a non-negative integer. 
\item
Because $\hnq_z$ can be both positive and negative, there is also the mirror image of the pole-skipping tower. Namely,  they start from $\hnq_z=-(s-1)$ and continue to $\hnq_z=-(s-1-n)$. For simplicity, we consider only the tower of item~2. 

\item 
For SAdS$_5$, the pole-skipping points depend on $q$. 
For the AdS soliton, the pole-skipping points depend on $\ki^2:=-\homega^2+q_x^2$.
\end{enumerate}
These results are easily expected, but we derive various field equations in the AdS soliton background explicitly and show that the above results indeed hold. Also, we have not discussed the boundary condition at the tip of the cigar $u=1$. One should ask the double Wick rotation including the boundary condition, so we discuss the boundary condition below.

%In order to distinguish the SAdS$_5$ and the AdS soliton, we use, \eg, $t'$, but we omit the prime in the rest of our paper. 
In order to distinguish the SAdS$_5$ and the AdS soliton, we use variables such as $\hatt$, but we omit ``~$\hat{~}$~" in the rest of our paper. 

%%------------------
\subsection{Boundary condition at the tip of the cigar}%\label{sec:}
%%------------------

%The boundary condition plays an important role, so let us discuss the boundary condition a little carefully. 
As we see below, all master fields behave as
\begin{align}
0 \sim Z''+ \frac{1}{u-1}Z'-\frac{\nq_z^2}{4(u-1)^2}Z~,
\quad(u\to1)~,
\label{eq:near_horizon}
\end{align}
near the tip of the cigar. 
Thus, there are two independent solutions:
\begin{align}
Z \propto (u-1)^{\pm\nq_z/2}~.
%\label{eq:}
%
\end{align}
%\begin{subequations}
%\label{eq:}
%\begin{align}
%
%\varphi &\sim (u-1)^{s_1}, \quad (u-1)^{s_2}, \\ 
%s_1 &=\frac{1+\nq_z}{2}~, \quad s_2=\frac{1-\nq_z}{2}~.
%\label{eq:}
%
%\end{align}
%\end{subequations}
For simplicity, we set $\nq_z>0$ below. The generic solution is a linear combination of these two solutions. The problem is how to choose the boundary condition at $u=1$. 

We follow the standard textbook treatment of quantum mechanics (Chap.~35 of \cite{LL}). By redefining $Z=G\varphi$ where $G=(u-1)^{-1/2}$, the master equation reduces to the \schrodinger\ problem near $u\to1$:
\begin{align}
0 &\sim \varphi''- \frac{\nq_z^2-1}{4x^2}\varphi~, 
\quad(x\to0)~, 
%V &\sim \frac{\nq_z^2-1}{4x^2}~,
%\label{eq:}
%
\end{align}
where $x:=u-1$.
%\begin{align}
%
%0= \varphi'' - \frac{\alpha^2}{x^2} \varphi~.
%\label{eq:}
%
%\end{align}
We impose the ``UV cutoff" at small $x$:
\begin{align}
V(x) = \left\{ 
\begin{array}{ll}
V(x)~, & x>x_0~,\\
V(x_0)=V_0~, & x<x_0~.
\end{array}
\right.
%\label{eq:}
%
\end{align}
As usual, we impose the conditions that the solution and its derivative are continuous at $x=x_0$, and we take the $x_0\to0$ limit. When $x>x_0$, the generic solution is
\begin{subequations}
%\label{eq:}
\begin{align}
\varphi &\sim C_1 x^{s_1}+C_2 x^{s_2}~, \\
s_1 &=\frac{1+\nq_z}{2}~, \quad s_2=\frac{1-\nq_z}{2}~,
%\label{eq:}
%
\end{align}
\end{subequations}
where %$s_1=(1+\nq_z)/2, s_2=(1-\nq_z)/2$ with 
$s_1>s_2$. One can show
\begin{align}
\frac{C_2}{C_1} \propto x_0^{s_1-s_2} \sim x_0^{\nq_z} \to 0~.
%\label{eq:}
%
\end{align}
So, it is enough to consider the solution $\varphi \sim (u-1)^{s_1}$ which falls faster. This is equivalent to choose $Z \sim (u-1)^{\nq_z/2}$ as the boundary condition. 

For a black hole, we impose the incoming-wave boundary condition $Z\sim (u-1)^{-i\nw/2}$, and the above choice is the analytic continuation from the black hole case. 
Both the geometry and the boundary condition are obtained by the double Wick rotation, so one expects that the pole-skipping results are also obtained by the double Wick rotation.

For simplicity, we set $\nq_z>0$ here, namely we set $\nq_z>0$ as the physical region. 
But $\nq_z<0$ should also be possible. In this case, one chooses $Z \sim (u-1)^{-\nq_z/2}$ as the boundary condition.

%%------------------
\subsection{Relation to quantum mechanical pole-skipping}%\label{sec:}
%%------------------

Ref.~\cite{Natsuume:2021fhn} finds a pole-skipping like phenomenon in quantum mechanics. It studies various potential problems with angular momentum $l=:\nu-1/2$:
\begin{subequations}
%\label{eq:}
\begin{align}
0 &= -\del_x^2\psi+V\psi-k^2\psi~, \\
V &= \frac{\nu^2-1/4}{x^2}+V_1~.
%\label{eq:}
%
\end{align}
\end{subequations}
It turns out that the $S$-matrix is not uniquely determined at
\begin{align}
\nu=-\frac{n}{2} \quad (n=1,2,\cdots)
%\label{eq:}
%
\end{align}
with appropriate $k$.
Ref.~\cite{Natsuume:2021fhn} considers
\begin{enumerate}
\item The Coulomb potential
\item The P\"{o}shl-Teller potentials
\end{enumerate}
as $V_1$. In those examples, near $x\to0$, the angular momentum part dominates:
\begin{align}
0 \sim -\del_x^2\psi + \frac{\nu^2-1/4}{x^2}~,
\quad(x\to0)~,
%\label{eq:}
%
\end{align}
so the solution is 
\begin{align}
\psi \sim x^{\lambda_\pm}~, \quad \lambda_\pm = \frac{1}{2}\pm\nu~.
%\label{eq:}
%
\end{align}
For physical momentum, $\nu\geq1/2$, so one chooses $\lambda_+$. Comparing with the AdS soliton case, $\nq_z=2\nu$, so this corresponds to choose $s_1$. The quantum mechanical pole-skipping occurs at $\nu=-n/2$ $(n=1,2,\cdots)$. This translates into the $\nq_z=-n$ pole-skipping for the AdS soliton. As far as the $x\to0$ behavior is concerned, the quantum mechanical pole-skip and the AdS soliton pole-skip reduce to the same problem.

%%%%%%%%%
\section{Pole-skipping in the AdS soliton geometry}\label{sec:pole_skip_soliton}
%%%%%%%%%

%%------------------
\subsection{Massive scalar}%\label{sec:}
%%------------------

We start with a massive scalar field. The field equation is given by
\begin{subequations}
%\label{eq:}
\begin{align}
0 &=(\nabla^2-m^2)\phi \\
&\propto 
\phi''+\left(\frac{f'}{f} - \frac{1}{u}\right)\phi' - \frac{ 4\nq_z^2+(4\nk^2+m^2/u) f }{4uf^2} \phi~.
%\label{eq:}
%
\end{align}
\end{subequations}
where $\nk^2:=-\nw^2+\nq_x^2$ with $\nw:=\omega/(2\pi/l), \nq_x:=q_x/(2\pi/l)$. Near the tip $u=1$, the field equation takes the form of \eq{near_horizon}:
\begin{align}
0 \sim \phi''+ \frac{1}{u-1}\phi'-\frac{\nq_z^2}{4(u-1)^2}\phi~.
%\label{eq:}
%
\end{align}
So, set the ansatz 
\begin{align}
\phi = (1-u^2)^{\nq_z/2}\Za~.
%\label{eq:}
%
\end{align}
Then, 
\begin{align}
0 \sim \Za''+ \frac{1+\nq_z}{u-1}\Za' + O\left(\frac{1}{u-1}\right)\Za~.
%\label{eq:}
%
\end{align}

Using the matrix formalism, one obtains the eigenvalue and the eigenvector $\veca_0$:
\begin{align}
\lambda = 0~, \quad
\veca_0
  = \begin{pmatrix} 
  1 \\
  b_1
    \end{pmatrix}
~, \quad 
b_1= -\frac{ 4\nk^2+6\nq_z^2+m^2 }{8(\nq_z+1)}~.
\end{align}
Thus, the first pole-skipping point is given by
\begin{align}
(\nq_z, \nk^2) = \left(-1, -\frac{6+m^2}{4} \right)~.
%\label{eq:}
%
\end{align}
The pole-skipping point lies outside the physical region $\nq_z \geq 0$. Now, move away from the pole-skipping point: $\nq_z=\nq_*+\delta\nq_z, \nk^2=\nk_*^2+\delta(\nk^2)$, where $(q_*,\nk_*^2)$ is the pole-skipping point. Then, $\veca_0$ actually has the slope-dependence:
\begin{align}
b_1 = \frac{3}{2} -\frac{\delta(\nk^2)}{2\delta\nq_z}~.
%\label{eq:}
%
\end{align}

At the next order, $b_2$ has a complicated form, so we give the $m=0$ result for simplicity:
\begin{align}
2b_2 = \frac{ 4\nk^4+3\nq_z^2(2+6\nq_z+3\nq_z^2)+4\nk^2(2+4\nq_z+3\nq_z^2) }{ 16(\nq_z+1)(\nq_z+2) }~,
%\label{eq:}
%
\end{align}
and $b_2$ is ambiguous at
\begin{align}
(\nq_z, \nk^2) = \left(-1, -\frac{3}{2} \right)~, \left(-1,\frac{1}{2} \right), (-2,-3\pm\sqrt{3})~.
%\label{eq:}
%
\end{align}
It includes the first pole-skipping point $(-1,-3/2)$. On the other hand, a new first pole-skipping point $(-1,1/2)$ seems to appear. But this is a fake one as discussed in \sect{matrix_formalism}.
In fact, it does not have the slope-dependence. Near the point, 
\begin{subequations}
%\label{eq:}
\begin{align}
\veca_1
  &= \begin{pmatrix} 
  b_1 \\
  2b_2
    \end{pmatrix}
~, \\ 
b_1&= \frac{ -2-\delta(\nk^2)+3\delta\nq_z }{2\delta\nq_z} \to \infty~,
2b_2= \frac{ 4\delta(\nk^2)+\delta\nq_z }{8\delta\nq_z}~.
%\label{eq:}
%
\end{align}
\end{subequations}

Finally, $b_3$ is ambiguous at 
%\begin{align}
%
%(\nq_z, \nk^2) = 
%&(-3,-3/2)~, (-3,(-15\pm2\sqrt{6})/2))~, \nonumber \\
%&(-2, -3\pm\sqrt{3})~, (-2,0)~, \nonumber \\
%& (-1, -3/2)~, (-1, (-3\pm\sqrt{3})/2)~.
%\label{eq:}
%
%\end{align}
% v2.1
\begin{align}
(\nq_z, \nk^2) = 
&(-3,-3/2)~, \left(-3,\frac{- 15 \pm 2\sqrt{6}}{2} \right)~, \nonumber \\
&(-2, -3\pm\sqrt{3})~, (-2,0)~, \nonumber \\
& (-1, -3/2)~, \left(-1, \frac{-3\pm\sqrt{3}}{2} \right)~.
%\label{eq:}
%
\end{align}
Note that the fake one from $b_2$ disappears. On the other hand, new fake ones appear: $(-2,0)~, (-1, (-3\pm\sqrt{3})/2)$.

%%------------------
\subsection{Gauge-invariant variables and master equations}\label{sec:master}
%%------------------

In order to discuss the Maxwell and gravitational perturbations, we first decompose the background spacetime into two parts $x^M=(x^a,y^i)$, where $x^a=(z,u)$ and $y^i=(t,x,y)$:
\be
ds_5^2 = g_{ab}(u) dx^a dx^b + \frac{1}{u} \eta_{ij}dy^i dy^j~.
%\label{eq:}
%
\ee
Here, the decomposition is chosen so that the $y^i$-part remains maximally symmetric. 
We decompose perturbations under the transformation of $y^i$. The scalar (vector) mode transforms as a scalar (vector) under the transformation.

As an example, consider the Maxwell perturbations $A_M=(A_a,A_i)$. One normally fixes the gauge $A_u=0$. We do not fix the gauge and carry out analysis in a fully gauge-invariant manner (\appen{gauge_inv_variables}). This is essentially the formalism by Kodama and Ishibashi \cite{Kodama:2003jz}.

For the Maxwell case, gauge-invariant variables are 
\begin{enumerate}
\item Scalar mode: $\mfA_z$ and $\mfA_u$.
\item Vector mode: $A_{T\,x}$ and $A_{T\,y}=A_y$.
\end{enumerate}

However, gauge-invariant variables are not independent, and they are related by the Maxwell equation. Both scalar variables obey first-order differential equations, so one gets a second-order differential equation for a single variable. They are referred to the master equation and the master field. 
 
Thus, there is only one degree of freedom, but the choice of a master field is not unique. One can choose any  linear combination of gauge-invariant variables as a master field. 
However, from the holographic point of view, it is natural to choose a master variable that does not involve $u$-derivatives of perturbations. This is because one imposes the Dirichlet boundary condition on the boundary. We choose such master fields below (see \appen{different} for more comments). The master equations we derive below coincide with the master equations for the SAdS$_5$ black hole  \cite{Kovtun:2005ev} after the double Wick rotation. 
%\footnote{Note that our gauge-invariant variables implicitly depend on $h'_{MN}$ through $\eta_a$ [see \eq{scalar-mfh-all}], so a general linear combination depends on $h'_{MN}$.}. 

%%------------------
\subsection{Maxwell field}%\label{sec:}
%%------------------

%%------------------
\subsubsection{Maxwell vector mode}%\label{sec:}
%%------------------

The vector mode $A_y$ is gauge invariant by itself. The Maxwell equation becomes
\begin{align}
0 &= A_y''+ \frac{f'}{f}A_y' - \frac{ \nq_z^2+\nk^2f }{uf^2} A_y~.
%\label{eq:}
%
\end{align}
Asymptotically, $A_y \sim A+B u$.
Near the tip $u=1$, the field equation takes the form of \eq{near_horizon}, so set the ansatz $A_y= (1-u^2)^{\nq_z/2}\Za$.
Then, 
\begin{align}
0 \sim \Za''+ \frac{1+\nq_z}{u-1}\Za' + O\left(\frac{1}{u-1}\right)\Za~.
%\label{eq:}
%
\end{align}

Using the matrix formalism, one obtains
\begin{align}
b_1= -\frac{ 2\nk^2+2\nq_z+3\nq_z^2}{4(\nq_z+1)}~.
\end{align}
Thus, the first pole-skipping point is given by
\begin{align}
(\nq_z, \nk^2) = \left(-1, -\frac{1}{2} \right)~.
%\label{eq:}
%
\end{align}
The pole-skipping point lies outside the physical region $\nq_z \geq 0$.

%%------------------
\subsubsection{Maxwell scalar mode}\label{sec:diffusive}
%%------------------

This mode corresponds to the ``diffusive mode" in the SAdS$_5$ case.
The gauge-invariant variables for the scalar mode are given by
\begin{subequations}
%\label{eq:}
\begin{align}
\mfA_z &= A_z - iq_z A_L~, 
\\ %\label{eq:inv_At} \\
\mfA_u &= A_u - A_L'~,
%\label{eq:inv_Ar}
%
\end{align}
\end{subequations}
(\appen{gauge_inv_variables}).
The Maxwell equation becomes
\begin{subequations}
%\label{eq:}
\begin{align}
0 &= \frac{i\nq_z}{2uf}\mfA_z + (f\mfA_u)'~,  \\
0 &= (2\nq_z^2+\nk^2f) \mfA_u + i\nq_z\mfA_z'~.
%\label{eq:}
%
\end{align}
\end{subequations}
We choose $\mfA_z$ as the master variable because $\mfA_u$ contains the $u$-derivative of the perturbation, $A_L'$. Then, the master equation is given by
\begin{align}
0 &= \mfA_z''+\frac{\nq_z^2f'}{(\nq_z^2+\nk^2f)f}\mfA_z' - \frac{ \nq_z^2+\nk^2f }{uf^2} \mfA_z~.
\label{eq:master_diffusive}
\end{align}
Asymptotically, $\mfA_z \sim A+B u$.
Near the tip $u=1$, the field equation takes the form of \eq{near_horizon}, so set the ansatz $\mfA_z= (1-u^2)^{\nq_z/2}\Za$.
Then, 
\begin{align}
0 \sim \Za''+ \frac{1+\nq_z}{u-1}\Za' + O\left(\frac{1}{u-1}\right)\Za~.
%\label{eq:}
%
\end{align}

Using the matrix formalism, one obtains
\begin{align}
b_1= -\frac{ 2\nk^2(\nq_z+2)+\nq_z^2(3\nq_z+2)}{4\nq_z(\nq_z+1)}~.
\end{align}
Thus, the first pole-skipping point is given by
\begin{align}
(\nq_z, \nk^2) = (0,0)~,\left(-1, \frac{1}{2} \right)~.
%\label{eq:}
%
\end{align}
The point $\nq_z=0$ corresponds to the hydrodynamic mode in the SAdS$_5$ case. 
While the point $\nq_z=-1$ lies outside the physical region $\nq_z \geq 0$, the $\nq_z=0$ point lies in the physical region.

%%------------------
\subsection{Gravitational field}%\label{sec:}
%%------------------

%%------------------
\subsubsection{Gravitational vector mode}%\label{sec:}
%%------------------

This mode corresponds to the ``shear mode" in the SAdS$_5$ case.
The gauge-invariant variables for the vector mode are given by
\begin{subequations}
%\label{eq:}
\begin{align}
\mfh_{zy} &= h_{zy}^{(1)} - iq_z h_y^{(1)}~, 
\\ %\label{eq:inv_At} \\
\mfh_{uy} &= h_{uy}^{(1)} - \frac{1}{u} (u h_y^{(1)})'~.
%\label{eq:inv_Ar}
%
\end{align}
\end{subequations}
The Einstein equation becomes
\begin{subequations}
%\label{eq:}
\begin{align}
0 &= \frac{i\nq_z}{2uf}\mfh_{zy}+(f\mfh_{uy})'~, \\
0 &= -\frac{2iu}{\nq_z} (\nq_z^2+\nk^2f) \mfh_{uy}+(u\mfh_{zy})'~.
%\label{eq:}
%
\end{align}
\end{subequations}
We choose $\mfh_{zy}$ as the master variable because $\mfh_{uy}$ contains the $u$-derivative of the perturbation. Then, the master equation is given by
\begin{align}
0 &= Z''
-\frac{(\nq_z^2+\nk^2f)f-\nq_z^2uf'}{uf(\nq_z^2+\nk^2f)} Z' 
-\frac{\nq_z^2+\nk^2f}{uf^2}Z~,
%+ \frac{-2\nq_z^2u^2+\nq_z^2f+k^2f^2}{uf (\nq_z^2+k^2 f)}Z' 
%- \frac{ \nq_z^4u+ k^2f^2(k^2u+f)+\nq_z^2f(1+2k^2u+u^2) }{uf(\nq_z^2+k^2f)} Z~,
%\label{eq:}
%
\end{align}
where $Z=u\mfh_{zy}$. 
Asymptotically, $Z \sim A+B u^2$.
Near the tip $u=1$, the field equation takes the form of \eq{near_horizon}, so set the ansatz $Z= (1-u^2)^{\nq_z/2}\Za$.
Then, 
\begin{align}
0 \sim \Za''+ \frac{1+\nq_z}{u-1}\Za' + O\left(\frac{1}{u-1}\right)\Za~.
%\label{eq:}
%
\end{align}

Using the matrix formalism, one obtains
\begin{align}
b_1= -\frac{ 2\nk^2(\nq_z+2)+3\nq_z^3}{4\nq_z(\nq_z+1)}~.
\end{align}
Thus, the first pole-skipping points are given by
\begin{align}
(\nq_z, \nk^2) = (0,0)~,\left(-1, \frac{3}{2} \right)~.
%\label{eq:}
%
\end{align}
The point $\nq_z=0$ corresponds to the hydrodynamic mode in the SAdS$_5$ case. 
While the point $\nq_z=-1$ lies outside the physical region $\nq_z \geq 0$, the $\nq_z=0$ point lies in the physical region.

%%------------------
\subsubsection{Gravitational scalar mode}%\label{sec:}
%%------------------

This mode corresponds to the ``sound mode" in the SAdS$_5$ case. The gauge-invariant variables are $\mfh_{zz}, \mfh_{zu}, \mfh_{uu}$, and $\mfh_{L}$. Their field equations are given in \eq{EOM_sound}.

From \eq{eta_a-all}, the gauge-invariant perturbations contain the parameters $\eta_u$, $\eta_z$, and their derivatives. $\eta_u$ contains a derivative of a metric. So, choose the combination of gauge-invariant variables which do not involve $\eta_u$. Such a master field is given by
\begin{subequations}
%\label{eq:}
\begin{align}
Z &= u \{ \mfh_{zz} - (f-uf') \mfh_L \} \\
&= u \biggl[ h_{zz}-2iq_zh_z -\biggl\{ q_z^2+ \frac{\ki_i^2}{3}(1+u^2) \biggr\} h_T 
\nonumber \\
& -(1+u^2)h_L \biggr]~.
%&= u \left[ h_{zz}-2iq_zh_z -\left\{ q_z^2+ \frac{\ki_i^2}{3}(1+u^2) \right\} h_T -(1+u^2)h_L \right]~.
%\label{eq:}
%
\end{align}
\end{subequations}
This is the master field, {\it e.g.}, by Kovtun and Starinets \cite{Kovtun:2005ev}.

One can obtain the master equation for $Z$ from $\mfh_{zz}$ and $\mfh_L$ equations.
The master equation is given by
\begin{align}
0 &= Z''- \frac{-3\nq_z^2(1+u^2)+\nk^2(-3+2u^2-3u^4)}{uf \{-3\nq_z^2+\nk^2(-3+u^2)\} }Z' 
\nonumber \\
+& \frac{3\nq_z^4+\nk^4(3-4u^2+u^4)+\nk^2\{ \nq_z^2(6-4u^2)-4u^3f \} }{ uf^2\{-3\nq_z^2+\nk^2(-3+u^2)\} } Z~.
%\label{eq:}
%
\end{align}
Asymptotically, $Z \sim A+B u^2$.
Near the tip $u=1$, the field equation takes the form of \eq{near_horizon}, so set the ansatz $Z= (1-u^2)^{\nq_z/2}\Za$.
Then, 
\begin{align}
0 \sim \Za''+ \frac{1+\nq_z}{u-1}\Za' + O\left(\frac{1}{u-1}\right)\Za~.
%\label{eq:}
%
\end{align}

Using the matrix formalism, one obtains
\begin{align}
b_1= -\frac{ 4\nk^4+9\nq_z^4+4\nk^2(3\nq_z^2+2\nq_z-2)}{4(\nq_z+1)(3\nq_z^2+2\nk^2)}~.
\end{align}
Thus, the first pole-skipping points are given by
\begin{align}
(\nq_z, \nk^2) = \left(1, -\frac{3}{2} \right)~, (0,0)~,\left(-1, \frac{1}{2}(1\pm2\sqrt{2}i) \right)~.
%\label{eq:}
%
\end{align}
The points $\nq_z=1$ and $\nq_z=0$ correspond to the chaotic mode and the hydrodynamic mode, respectively, in the SAdS$_5$ case. 
In the matrix formalism, one obtains all pole-skipping points including the ``chaotic" and the ``hydrodynamic" points as promised.
While the point $\nq_z=-1$ lies outside the physical region $\nq_z \geq 0$, the $\nq_z=1,0$ points lie in the physical region.

%%%%%%%%%
\section{Discussion}%\label{sec:}
%%%%%%%%%

%%------------------
\subsection{Physical implication}%\label{sec:}
%%------------------
In this paper, we study the pole-skipping in the AdS soliton geometry. Even though the AdS soliton is not a black hole, the field equations have regular singularities at the tip of the cigar, so the pole-skipping occurs. 

It is interesting to explore the physical implications. However, even in the \bh case, the physical implications of the pole-skipping are unclear in general. 
This is because many pole-skipping points do not lie in the physical region (the wave number $q$ is complex in general). This makes the physical interpretations difficult in general. The exceptions are the chaotic and the hydrodynamic pole-skippings. 

In the AdS soliton case, the pole-skipping points in general do not lie in the physical region at large $n$. 
However, the ``chaotic" and ``hydrodynamic" pole-skipping points lie in the physical region, 
so it is interesting to explore physical implications. 

%Note that even  in the \bh case, many pole-skipping points do not lie in the physical region because the wave number $q$ is complex in general. This makes the physical implications of the pole-skipping unclear in general.
%\footnote{Even in the \bh case, many pole-skipping points lie outside the physical region. In the \bh case, the physical region is $\omega$ in the lower-half plane and real $q$. This would exclude the chaotic pole-skipping at $\nw=+i$, so it is subtle to exclude pole-skipping points outside the physical region. For the pole-skipping points $\nw=-i$, (1) the Maxwell scalar and the gravitational vector modes lie in the physical region (real $q$), but (2) the Maxwell vector, the gravitational tensor, and gravitational scalar modes do not lie in the physical region.}

The pole-skipping itself occurs even in the AdS soliton, but the physical interpretation is different from the \bh case. First, for a black hole, the chaotic pole-skipping point lies in the upper-half $\omega$-plane, which suggests a chaotic behavior. But the AdS soliton is not a black hole, and one does not expect the chaotic behavior. In fact, both $\nq_z$ and $\nk^2$ are real there.

% v2
Second, we make $\nq_z$ continuous following the conventional pole-skipping analysis, but  $\nq_z$ is physically discrete $\nq_z\in \mathbb{Z}$. One may wonder if the pole-skipping has any physical relevance for the AdS soliton. 
%It is interesting to explore the issue as well (see, \eg, \cite{second}).

Actually, there is an evident physical interpretation for the ``chaotic" pole-skipping in the AdS soliton. The chaotic pole-skipping point is located at $(\nq_z,\nk^2)=(1,-3/2)$. This is a pole-skipping point, so 
%(1) it would be a pole, but (2) the residue of the pole vanishes. 
\begin{enumerate}
\item It would be a pole. 
\item But the residue of the pole vanishes. 
\end{enumerate}
The former implies that the dual field theory would have a normal mode with $(\nq_z,m_3^2)=(1,6)$, where $m_3^2=-p^2=-4\nk^2$ is the dual $(2+1)$-dimensional mass. However, the latter implies that the state is actually \textit{missing} due to the pole-skipping.

In the \bh case, near a pole-skipping point, the Green's function 
typically takes the form 
\begin{align}
G^R \propto \frac{\delta\omega+\delta q}{\delta\omega-\delta q}~,
%\label{eq:}
%
\end{align}
so it depends on the slope $\delta q/\delta \omega$ how one approaches the pole-skipping point. However, if one first fixes $\delta q=0$, one gets $G^R =(\text{constant})$, so the pole would disappear.

This is the situation that happens in the AdS soliton case. 
 In the AdS soliton case, $\nq_z$ is actually discrete, and one first fixes $\nq_z=1$, and so on, so one cannot choose the slope. Instead, the pole-skipping appears as the ``missing state."

Even though the black hole and the AdS soliton are related by a double Wick rotation, the physical interpretations of the ``chaotic" pole-skipping are very different. The black hole and the AdS soliton have very different physical interpretations in general. For the ``chaotic" pole-skipping, 
\begin{enumerate}
\item
The black hole case implies a chaotic behavior.
\item
The AdS soliton case implies a missing state. 
\end{enumerate}
%We do not see any clear connection. However, note that the chaotic pole-skipping $\nw_{+1}=+i$ does not lie in the physical region (one would take the lower-half $\omega$-plane as the physical region). Because $i\hat{q}_z=\omega$, the ``chaotic" pole-skipping in the AdS soliton now lies in the physical region ($q_z>0$), so it affects the spectrum.

We will explore this issue further in a separate paper, in particular the mass spectrum in details \cite{Natsuume:2023hsz}.

%%------------------
\subsection{Other non-black-hole backgrounds}%\label{sec:}
%%------------------
We focus on the AdS soliton, but a similar analysis should be possible for the other non-\bh geometries with $S^1$. The Witten geometry is an example \cite{Witten:1998zw}. This geometry is obtained by the double Wick rotation from the D4-brane geometry.
%From the gauge theory point of view, a confining gauge theory with $S^1$ typically has the first-order phase transition, the Hawking-Page transition. But there are many confining gauge theories other than the first-order phase transition, and their dual geometries do not have $S^1$. 

Of course, there are many non-black-hole geometries without $S^1$. For a non-black-hole background, field equations in general would not have regular singularities (except $u=\infty$) and have only ordinary points. However, as mentioned in \sect{matrix_formalism}, there are 2 kinds of pole-skipping. In the \bh case, the pole-skipping in the lower-half $\omega$-plane comes from regular singularities. But the pole-skipping in the upper-half $\omega$-plane has a different origin. In this case, the coefficients of field equations themselves partly have the structure $0/0$. In particular, the hydrodynamic mode should survive as a pole-skipping point. 

As a simple example, consider the cutoff SAdS where we impose the IR cutoff at $u=u_0<1$. Because the cutoff is an ordinary point, one can expand the solution as a Taylor series:
\begin{subequations}
%\label{eq:}
\begin{align}
0 &= Z''+PZ'+QZ~, \\
P &= \sum_{n=0} P_n(u-u_0)^n~, \\
Q &= \sum_{n=0} Q_n(u-u_0)^n~, \\
Z &= \sum_{n=0} a_n(u-u_0)^n~.
%
%\label{eq:}
%
\end{align}
\end{subequations}
At the lowest order, one gets
\begin{align}
0= Q_0a_0+P_0a_1+2a_2~.
\label{eq:eom_regular}
\end{align}
One needs to impose a boundary condition at the cutoff. If one imposes the Dirichlet boundary condition, $a_0=0$. Then, \eq{eom_regular} determines $a_2$ from $a_1$. However, if $P_0=0/0$, both $a_1$ and $a_2$ are free parameters, and the solution is not uniquely determined. For example, for the Maxwell scalar mode, $P_0=0/0$ at $\omega=q=0$ [see \eq{master_diffusive} for the AdS soliton counterpart]. 

It is unclear if there are any other pole-skipping points for non-\bh geometries. It would be interesting to explore the issue further.

%%%%%%%%%
\section*{Acknowledgments}%\label{sec:}
%%%%%%%%%

%%%%%%%%%
%\begin{acknowledgments}

%MN would like to thank ...
This research was supported in part by a Grant-in-Aid for Scientific Research (17K05427) from the Ministry of Education, Culture, Sports, Science and Technology, Japan. 

\appendix

%\small

%%%%%%%%%
\section{The SAdS$_5$ black hole master equations}\label{sec:KS}
%%%%%%%%%

Below we summarize the master equations for the SAdS$_5$ black hole 
% v2.1
for the reader's convenience \cite{Kovtun:2005ev}:

\begin{enumerate}
\item
Maxwell vector mode:
\begin{align}
0 &= Z''+ \frac{f'}{f}Z' + \frac{ \nw^2-\nq^2f }{uf^2} Z~,
%\label{eq:}
%
\end{align}
where $'=\del_u$ and $\nw:=\omega/(2\pi T), \nq:=q/(2\pi T)$.

\item
Maxwell scalar mode:
\begin{align}
%
%corrected
0 &= Z''+\frac{\nw^2f'}{(\nw^2-\nq^2f)f}Z' + \frac{ \nw^2-\nq^2f }{uf^2} Z~.
%\label{eq:}
%
\end{align}

\item
Gravitational tensor mode (massless scalar):
\begin{align}
0 &= Z''+ \left(\frac{f'}{f} - \frac{1}{u}\right)Z' + \frac{ \nw^2-\nq^2f }{uf^2} Z~.
%\label{eq:}
%
\end{align}

\item
Gravitational vector mode:
\begin{align}
0 &= Z''
-\frac{(\nw^2-\nq^2f)f - \nw^2uf'}{uf(\nw^2-\nq^2f)} Z' 
+\frac{\nw^2-\nq^2f}{uf^2}Z~.
%\label{eq:}
%
\end{align}

\item
Gravitational scalar mode:
\begin{align}
0 &= Z''- \frac{3\nw^2(1+u^2)+\nq^2(-3+2u^2-3u^4)}{uf \{3\nw^2+\nq^2(-3+u^2)\} }Z' 
 \\
+& \frac{3\nw^4+\nq^4(3-4u^2+u^4)-\nq^2\{ \nw^2(6-4u^2)+4u^3f \} }{ uf^2\{3\nw^2+\nq^2(-3+u^2)\} } Z~.
\nonumber
%\label{eq:}
%
\end{align}
\end{enumerate}

All master fields behave as
\begin{align}
0 \sim Z''+ \frac{1}{u-1}Z'+\frac{\nw^2}{4(u-1)^2}Z~,
\quad(u\to1)~,
%\label{eq:}
%
\end{align}
near the horizon, and one imposes the incoming-wave boundary condition:
\begin{align}
Z \propto (u-1)^{-i\nw/2}~.
%\label{eq:}
%
\end{align}

%%%%%%%%%
\section{Properties of the AdS soliton}\label{sec:properties}
%%%%%%%%%

%%------------------
\subsection{Phase transition}%\label{sec:}
%%------------------

At high temperature, the AdS soliton undergoes a first-order phase transition (the Hawking-Page transition) to the SAdS black hole. This describes the confinement/deconfinement transition in the dual gauge theory.

Because the (uncompactified) SAdS \bh is scale invariant, the only dimensionful quantity is $T$. Thus, the free energy should take the form
\begin{align}
F_\text{BH} = -c T^4 V_3~,
%\label{eq:}
%
\end{align}
where 
% v2.1
$c=\pi^4/(16\pi G_5)=\pi^2N_c^2/8$, 
%$c=r_0^4/(16\pi G_5)=N_c^2/(8\pi)^2$, 
and $V_3$ is the gauge theory volume. The $S^1$-periodicity $l$ appears only in $V_3$. The AdS soliton has the same Euclidean geometry, so the free energy takes the same form when expressed in terms of the ``horizon" radius $r_0$. But, for the AdS soliton, $r_0$ is related to the $S^1$ periodicity $l$, so 
\begin{align}
F_\text{soliton} = -\frac{c}{l^4}V_3~.
%\label{eq:}
%
\end{align}
Then, the free energy difference is 
\begin{align}
\Delta F = F_\text{BH} - F_\text{soliton} 
\propto - \left(T^4 - \frac{1}{l^4}\right) V_3~.
%\label{eq:}
%
\end{align}
So, at low temperature $T<1/l$, the stable solution is the AdS soliton. At high temperature $T>1/l$, the stable solution is the black hole. Because the \bh has $O(N_c^2)$ entropy, the entropy is discontinuous at $Tl=1$. The first derivative of free energy, $S = -\del_T F$, is discontinuous there, so this is a first-order phase transition. 

%%------------------
\subsection{Energy-momentum tensor}%\label{sec:}
%%------------------

The SAdS$_5$ \bh has the following energy-momentum tensor:
\begin{align}
T_{\mu\nu} 
= \frac{r_0^4}{16\pi G_5} \text{diag}(3,1,1,1)
= \frac{\pi^2}{8} N_c^2T^4 \text{diag}(3,1,1,1)~.
%\label{eq:}
%
\end{align}
The energy-momentum tensor for the AdS soliton is computed in Ref.~\cite{Myers:1999psa}. One can obtain it by the double Wick rotation from the SAdS result:
\begin{align}
T_{\mu\nu} = \frac{\pi^2}{8} \frac{N_c^2}{l^4} \text{diag}(-1,1,1,-3)~.
%\label{eq:}
%
\end{align}
It remains traceless. The nonvanishing energy density is interpreted as the Casimir energy. The directions $(t,x,y)$ 
% v2.1
have the Lorentz invariance, so $T_{ij} (i=t,x,y)$ has the Lorentz invariance and is proportional to $\eta_{ij}$. The Lorentz invariance is broken in the $z$-direction, so one obtains an anisotropic pressure. 

%%------------------
\subsection{Mass gap}\label{sec:mass_gap}
%%------------------

For the SAdS black hole, one imposes the incoming-wave boundary condition at the horizon. Because a perturbation is absorbed by the black hole, one obtains quasinormal modes, namely poles are located in the complex $\omega$-plane. The AdS soliton does not have a horizon, and the geometry smoothly ends at $u=1$, so one obtains normal modes. 

For simplicity, consider the massless scalar field $0=\nabla^2\phi$, and consider the perturbation of the form $e^{i\ki_ix^i}$. The field has the asymptotic behavior 
\begin{align}
\phi \sim A+Bu^2~, \quad (u\to0)~,
%\label{eq:}
%
\end{align}
and the Green's function is given by
\begin{align}
G^R \propto \frac{B}{A}~.
%\label{eq:}
%
\end{align}
So, a pole corresponds to $A=0$. Then, it is enough to solve the perturbation equation under the boundary condition $A=0$. 

Near the tip of cigar $u=1$, the field equation becomes
\begin{align}
0 \sim \phi''+\frac{1}{u-1}\phi'~,
\quad(u\to1)~,
%\label{eq:}
%
\end{align}
when $q_z=0$, so the solution is 
\begin{align}
\phi \sim C_1+C_2 \ln(1-u)~.
%\label{eq:}
%
\end{align}
We impose $C_2=0$ from the regularity condition at $u=1$.

One can solve the perturbation equation by a power series expansion around the tip of the cigar $u=1$:
\begin{align}
\phi= \sum_{n=0}^\infty a_n(u-1)^{n+\lambda}~.
%\label{eq:}
%
\end{align}
Substituting this into the field equation, one gets the indicial equation at the lowest order: $\lambda^2=0$. The boundary condition $A=0$ corresponds to
\begin{align}
\phi|_{u=0} = \sum_{n=0}^N (-)^n a_n =0~.
\label{eq:bc_infty}
\end{align}
One truncates the series after a large number of terms $n=N$. One can check the accuracy as one goes to higher series. The problem has a nontrivial solution only for particular values of $\ki^2=-m_3^2$ which give the mass spectrum. The first few states are $m_3^2\approx 11.59, 34.53, \cdots$.

%%%%%%%%%
\section{Gauge-invariant variables}\label{sec:gauge_inv_variables}
%%%%%%%%%

We follow Ref.~\cite{Natsuume:2019sfp} with slight changes in the conventions.
First, decompose the background spacetime into a two-dimensional space $x^a$ and a $p$-dimensional spacetime $y^i$:
% v2.1
\be
ds_{p+2}^2 = g_{ab}(x) dx^a dx^b +e^{2\vphi} g_{ij}(y)dy^i dy^j~.
%\label{eq:}
%
\ee
Here, the decomposition is chosen so that the $y^i$-part remains maximally symmetric. 
\begin{enumerate}
\item
For the SAdS$_5$, the metric is \eq{sads5}, so $x^a=(t,u)$, $y^i=(x,y,z)$ and $g_{ij}=\delta_{ij}$.
\item
For the AdS soliton, the metric is \eq{ads_soliton}, so $x^a=(z,u)$, $y^i=(t,x,y)$ and $g_{ij}=\eta_{ij}$.
\end{enumerate}
Below, we focus on the AdS soliton case. 
$e^{2\vphi}=1/u$, and 
\begin{align}
  & g_{ab}
  = \begin{pmatrix} g_{zz}~ & 0 \\ 0 & g_{uu} \end{pmatrix}
~.
\label{A-eq:bg_metric} 
\end{align}
We decompose perturbations under the transformation of $y^i$. The scalar (vector) mode transforms as a scalar (vector) under the transformation. We consider the perturbation of the form
\begin{align}
\phi(u) e^{i\ki_i x^i+iq_zz}~,
%\label{eq:}
%
\end{align}
where $\ki_i=(-\omega, q_x,0)$.%
\footnote{
For the quantities defined in the $p$-dimensional subspacetime such as $\ki^i$, the index $i$ is raised and lowered with $\eta_{ij}$, \ie, $\ki^i=\eta^{ij}\ki_j$. 
%For simplicity, we consider the $p$-dimensional metric which is proportional to $\delta_{ij}$, but the extention to $\gamma_{ij}(x)$ is easy. Replace $\del_i$ with $\bmD_i$, the covariant derivative with respect to $\gamma_{ij}$. Some expressions must be symmetrized since $\bmD_i$ do not commute.
}

%%------------------
\subsection{Maxwell perturbations}%\label{sec:}
%%------------------

%For simplicity, we assume that the background spacetime is 4-dimensional Minkowski spacetime. Write $A_M=(A_a, A_i)$, where $x^a=(t,r)$ and $x^i=(x,y)$. 
The Maxwell perturbations consist of $A_M=(A_a, A_i)$. $A_i$ can be decomposed as 
\begin{align}
A_i = \del_iA_L+A_{T\,i}~, \quad
\del^i A_{T\,i} =0~,
\label{eq:dec-Ai}
\end{align}
or 
\begin{align}
A_i = i\ki_i A_L+A_{T\,i}~, \quad
\ki^i A_{T\,i} =0~.
%\label{eq:}
%
\end{align}
Thus, for $p=3$,
\begin{enumerate}
\item
The scalar mode consists of 3 perturbations $A_a (A_z, A_u)$ and $A_L$. 
\item
One can use the Lorenz gauge condition to eliminate a component of $A_{T\,i}$, \eg, $A_{T\,t}$. Then, the vector mode consists of 2 perturbations $A_{T\,x}$ and $A_{T\,y}=A_y$.  They satisfy the identical equation. 
\end{enumerate}
In total, there are 5 perturbations.

The scalar mode has 3 perturbations, but one is redundant due to the gauge symmetry. The gauge transformation $\delta A_M = -\del_M \lambda$ becomes
\begin{subequations}
%\label{eq:}
\begin{align}
\delta A_a &= -\del_a \lambda~, 
\label{eq:del_Aa} \\
\delta A_i &= i\ki_i \delta A_L + \delta A_{T\,i} = -i\ki_i \lambda
\label{eq:del_Ax}
\end{align}
\end{subequations}
so that 
%\margin{?}
\begin{align}
\delta A_{T\,i} =0~, \quad 
\delta A_L=-\lambda~.
\label{eq:del_AL}
\end{align}
The gauge-invariant variables are obtained by eliminating the gauge parameter $\lambda$. The variables $A_{T\,i}$ are gauge invariant by themselves. From \eq{del_AL}, $\lambda$ is expressed by $A_L$ as $\lambda = -\delta A_L$. Substituting this into \eq{del_Aa} gives
\be
\delta(A_a - \del_a A_L) = 0~,
%\label{eq:}
%
\ee
so the gauge-invariant scalar perturbations are given by
\begin{align}
\mfA_a := A_a - \del_a A_L~.
\label{eq:inv_Aa}
\end{align}

%%------------------
\subsection{Gravitational perturbations}%\label{sec:}
%%------------------

The gravitational perturbations consist of $h_{MN} = (h_{ab}, h_{ai}, h_{ij})$. Again, perturbations are decomposed 
%v 2.1
as scalar, vector, and tensor modes. $h_{ab}$ gives 3 scalar perturbations. Just like the Maxwell perturbations, $h_{ai}$ is decomposed as
\begin{align}
h_{ai} = \del_i h_a + h^{(1)}_{ai}~, \quad
\del^i h^{(1)}_{ai} =0~.
\label{eq:dec-hai}
\end{align}
$h_a$ gives 2 scalar perturbations, and $h^{(1)}_{ai}$ gives $2(p-1)$ vector perturbations. The superscript ``$(1)$" refers to the number of index $i$ (``spin"). Similarly, $h_{ij}$ is decomposed as
\begin{align}
h_{ij} =: h_L\, \eta_{ij} + \calP_{ij}\,h_T
  + 2 \del_{(i} h^{(1)}_{T\, j)} + h^{(2)}_{T\, ij}~,
\label{eq:dec-hij}
\end{align}
where
\begin{align}
  & %\del^i h^{(1)}_{ai} = 
  \del^i h^{(1)}_{T\, i} = 0~,
& & \del^j h^{(2)}_{T\, ij} = 0~,
& & h^{(2)}_{T\, i}{}^i = 0~,
%\label{eq:}
%
\end{align}
and $\calP_{ij}$ is the projection operator given by
\begin{align}
\calP_{ij} := \del_i \del_j - \frac{1}{p} \eta_{ij}\, \del_k^2~.
%\label{eq:}
%
\end{align}
The first term of $h_{ij}$ ($h_L$) is the trace part which is a scalar perturbation. The rest is the traceless part which is decomposed as a scalar $h_T$, vector $h^{(1)}_{T\, j}$, and tensor perturbations $h^{(2)}_{T\, ij}$. Thus, 
\begin{enumerate}
\item The scalar mode consists of 7 perturbations $(h_{ab}, h_a, h_L, h_T)$. 
\item The vector mode consists of perturbations $(h^{(1)}_{ai}, h^{(1)}_{T\, i})$. The former has $2(p-1)$ components, and the latter has $(p-1)$ components, so there are $3(p-1)$ components. 
\item The tensor mode consists of $h^{(2)}_{T\, ij}$ which has $(p+1)(p-2)/2$ components. 
\end{enumerate}
In total, there are $(p+2)(p+3)/2$ components.

Again consider the gauge transformation $\delta x^M = - \xi^M$. %!
The infinitesimal transformation $\xi^i$ is decomposed as
\begin{align}
\xi_i =: \del_i\xi_L+\xi_{T\,i}~, \quad
\del^i \xi_{T\,i} =0~.
\label{eq:dec-xi}
\end{align}
The scalar part has 3 components, $\xi_a$ and $\xi_L$, and the vector part has $(p-1)$ components, $ \xi_{T\,i}$. 

%%------------------
\subsubsection{Tensor mode}%\label{sec:}
%%------------------

% v2.1
The tensor mode is gauge invariant by itself:
\begin{align}
\delta h^{(2)}_{T\, ij}=0~.
%\label{eq:}
%
\end{align}
The combination $(uh^{(2)}_{T\, ij})$ satisfies the field equation for the minimally-coupled massless scalar field, so we do not discuss this mode further.

%%------------------
\subsubsection{Vector mode}%\label{sec:}
%%------------------

Under the gauge transformation $\delta x^M =-\xi^M$, the vector mode transforms as %!
\begin{subequations}
%\label{eq:}
\begin{align}
  & \delta h^{(1)}_{ai}
  = \partial_a \xi_{T\, i}
       - 2\, \xi_{T\, i} \big( \partial_a \vphi \big)~,
\label{eq:del_hai} \\
  & \delta h^{(1)}_{T\, i} = \xi_{T\, i}~.
\label{eq:del_hTi} 
\end{align}
\end{subequations}

In order to obtain gauge-invariant variables, we again express gauge parameters $\xi_{T\, i}$ by perturbations. \eq{del_hTi} expresses $\xi_{T\, i}$ by $\delta h^{(1)}_{T\, i}$. Substituting \eq{del_hTi} into \eq{del_hai} gives
% v2.1
\begin{align}
\delta \left( h^{(1)}_{ai} - \partial_a h^{(1)}_{T\, i}
  + 2\, h^{(1)}_{T\, i}~\partial_a \vphi  \right) = 0~,
%\quad\to\quad
%  \mfh_{ai}
%  &:= h^{(1)}_{ai} - e^{2 \vphi}\,
%    \partial_a \Big( e^{- 2 \vphi}\, h^{(1)}_{T\, i} \Big)~.
%\label{eq:}
%
\end{align}
so the gauge-invariant vector perturbations are
\begin{align}
 \mfh_{ai}
  &:= h^{(1)}_{ai} - e^{2 \vphi}\,
    \partial_a \Big( e^{- 2 \vphi}\, h^{(1)}_{T\, i} \Big)~.
%\label{eq:}
%
\end{align}
Eliminating $\xi_{T\,i}$ gives $2(p-1)$ gauge-invariant perturbations. 

%%------------------
\subsubsection{Scalar mode}%\label{sec:}
%%------------------

\paragraph{The gauge-invariant variables:}
Under the gauge transformation, the scalar mode transforms as
\begin{subequations}
\label{A-eq:del_G-h_M-all}
\begin{align}
  & \delta h_{ab} = 2\, \twonabla_{(a} \xi_{b)}~,
\label{eq:del_hab} \\
  & \delta h_{a}
  = \xi_a + \partial_a \xi_L
  - 2\, \xi_L \big( \partial_a \vphi \big)~,
\label{eq:del_ha} \\
  & \delta h_L
  = \xi^a\, \twonabla_a e^{2 \vphi}
  + \frac{2}{p}\, \del_k^2\, \xi_L~,
\label{eq:del_hL} \\
  & \delta h_T = 2\, \xi_L~,
\label{eq:del_hT} 
\end{align}
\end{subequations}
where $\twonabla_a$ is the covariant derivative with respect to $g_{ab}$. 

\eq{del_hT} expresses $\xi_L$ by $\delta h_T$. Substituting \eq{del_hT} into \eq{del_ha}, $\xi_a$ is expressed by $\delta h_a$ and $\delta h_T$:
\begin{subequations}
\label{A-eq:xi_a-all}
\begin{align}
  & \xi_a = -\delta \eta_a %!
  ~,
\label{A-eq:xi_a} \\
  & \eta_a
  := \frac{1}{2}\, \partial_a h_T - h_T \partial_a \vphi - h_{a}~.
%  = \frac{ e^{2 \vphi} }{2}\,
%    \partial_a \Big( e^{- 2 \vphi}\, h^{(0)}_T \Big) - h_{a}
\label{eq:eta_a}
\end{align}
\end{subequations}
Substituting $\xi_a$ into \eq{del_hab}, one obtains
\begin{align}
\delta \left(h_{ab} + 2\, \twonabla_{(a} \eta_{b)} \right) =0~, %\quad\to\quad
%   \mfh_{ab}
%  &:= h_{ab} + 2\, \twonabla_{(a} \eta_{b)}~.
%\label{eq:inv_hab} 
%
\end{align}
so one gets the gauge-invariant perturbations
\begin{align}
\mfh_{ab} &:= h_{ab} + 2\, \twonabla_{(a} \eta_{b)}~.
%\label{eq:}
%
\end{align}

Similarly, \eq{del_hL} becomes
\begin{align}
\delta \left( h_L + \eta^a\, \twonabla_a e^{2 \vphi}
  - \frac{1}{p}\, \del_i^2\, h_T \right) = 0~,
%\quad\to\quad
%   \mfh_L
%  &:= h_L + \eta^a\, \twonabla_a e^{2 \vphi}
%  - \frac{1}{p}\, \del_i^2\, h_T~.
%\label{eq:inv_hL} 
%
\end{align}
which gives
\begin{align}
\mfh_L &:= h_L + \eta^a\, \twonabla_a e^{2 \vphi} - \frac{1}{p}\, \del_i^2\, h_T~.
\label{eq:inv_hL} 
%\label{eq:}
%
\end{align}
The scalar mode has 7 perturbations, but 4 gauge-invariant perturbations remain after one eliminates $\xi_L$ and $\xi_a$.

Writing these formulae in components for $p=3$, one gets
%For $p=2$, gauge-invariant vector perturbations are 
%\begin{align}
%
%\mfh_{vy} &= h_{vy} +\frac{\omega}{q} h_{xy}~, \\
%\mfh_{ry} &= h_{ry} -\frac{r^2}{iq}\left( \frac{ h_{xy} }{r^2} \right)'~.
%\label{eq:}
%
%\end{align}
%Gauge-invariant scalar perturbations are 
%
\begin{subequations}
\label{eq:scalar-mfh-all}
\begin{align}
\mfh_{zz} & = h_{zz} + 2iq_z \eta_z + \frac{g_{zz}'}{g_{uu}}\eta_u~,
\label{A-eq:mfh_vv-I} \\
\mfh_{zu} & = h_{zu} + iq_z \eta_u - \frac{g_{zz}'}{g_{zz}}\eta_z+\eta_z'~,
%  + \eta_v'
%  - \left( i\, \omega + \bmg_{vv}' \right)\, \eta_r
\label{A-eq:mfh_vr-I} \\
\mfh_{uu} & = h_{uu} - \frac{g_{uu}'}{g_{uu}}\eta_u + 2\eta_u'~,
%  &:= h_{rr} + 2\, \eta_r'
\label{A-eq:mfh_rr} \\
\mfh_L & = h_L +\frac{\ki_i^2}{3} h_T - \frac{1}{u^2g_{uu}} \eta_u~.
 % = h_{yy}
 % + 2r ( \eta_v - \bmg_{vv} \eta_r)~.
\label{A-eq:mfh_L-I}
\end{align}
\end{subequations}
From \eq{eta_a}, $\eta_a$ becomes 
\begin{subequations}
\label{eq:eta_a-all}
\begin{align}
\eta_z &= \half iq_z h_T - h_z~,
%   \eta_v
%  &  = - \frac{1}{i\, q} \left( h_{vx}
%    + \frac{\omega}{q}\, \frac{ h_{xx} - h_{yy} }{2}
%  \right)
\label{A-eq:eta_v} \\
\eta_u &= \frac{1}{2u}(uh_T)'-h_u~.
%   \eta_r
%  &  = \frac{1}{i\, q}\, \left\{ \frac{ r^2 }{i\, q}\,
%       \left( \frac{ h_{xx} - h_{yy} }{2\, r^2} \right)'
%    - h_{rx} \right\}
%  ~.
\label{A-eq:eta_r}
\end{align}
\end{subequations}

\paragraph{The field equations:}
The scalar mode has 4 variables $\mfh_{zz},\mfh_{zu}, \mfh_{uu}, \mfh_{L}$. 
The linearized Einstein equation reduces to 
\begin{subequations}
\label{eq:EOM_sound}
\begin{align}
0&= \mfh_{zz}+ f \mfh_L + 4uf^2 \mfh_{uu}~, 
\label{eq:EOM_const0} \\
0&= \frac{2\nk^2+3u}{2f}\mfh_{zz} + \{3\nq_z^2+\nk^2(2-f)\}\frac{\mfh_{zu}}{i\nq_z} 
\nonumber \\
&+ \frac{6\nq_z^2+4\nk^2f-3u(2-f)}{2f} \mfh_L~, 
\label{eq:EOM_const1} \\
0&= \left( u\frac{d}{du}+\frac{1}{f} \right)\mfh_{zz} -2(1-f)\mfh_L 
\nonumber \\
&+ 2 \frac{3\nq_z^2+2\nk^2f}{3}\frac{u \mfh_{zu}}{i\nq_z} +2uf(2-f)\mfh_{uu}~, 
\label{eq:EOM_zz} \\
0&= \left( u\frac{d}{du}+\frac{1}{f} \right)\mfh_L - \frac{2\nk^2}{3}\frac{u\mfh_{zu}}{i\nq_z} + 2uf\mfh_{uu}~.
\label{eq:EOM_L} 
%\\
%0&= \left( u\frac{d}{du}+1-\frac{2}{f} \right)\frac{u\mfh_{zu}}{i\nq_z} - \frac{u}{f} \mfh_L - 2u^2 \mfh_{uu}~.
%\label{eq:EOM_zu}
%
\end{align}
\end{subequations}
\pagebreak
Here, we arranged field equations. Namely, these equations are not just the bare Einstein equation components. The bare equations involve first and second derivatives in $u$. 
% v2.1
Combining the equations appropriately, one can eliminate all second derivatives. The resulting equations involve various first derivatives. Combining equations further 
% v2.1
eliminates some first derivatives, and one obtains equations where the first derivative appears at most once in each equation. Finally, we use the constraint equations for cosmetic purposes.

There are 2 constraint equations without $u$-derivatives and 2 differential equations with one $u$-derivative. 
%The latter 3 are not independent: one is redundant from the constraint equation \eqref{eq:EOM_const1}. 
The constraint equations \eqref{eq:EOM_const0} and \eqref{eq:EOM_const1} allow us to choose 2 independent variables. Both obey first-order differential equations, so one gets a second-order differential equation for a single variable.

%%%%%%%%%
\section{The choice of master field}\label{sec:different}
%%%%%%%%%

As mentioned in \sect{master}, the choice of a master field is not unique. One needs to find which variable is most suitable, or one needs to take all variables into account \cite{Natsuume:2019sfp}. This is sometimes problematic for the pole-skipping analysis. One cannot find some pole-skipping points if one chooses a different variable. In this Appendix, we show this explicitly for the Maxwell scalar mode and for the gravitational vector mode.

For the Maxwell scalar mode, there are 2 gauge-invariant variables $\mfA_z, \mfA_u$, and we choose $\mfA_z$. If one chooses $\mfA_u$ as the master variable, the master equation is given by
\begin{align}
0= Z_2''+\left( \frac{f'}{f}+\frac{1}{u} \right) Z_2'-\frac{\nq_z^2+\nk^2f}{uf^2}Z_2~, 
%\label{eq:}
%
\end{align}
where $Z_2=f\mfA_u$. 
%Asymptotically, $Z\sim u^\lambda$ with $\lambda=0$ (degenerate? $\lambda$ should differ by 1?).In
In this case, one obtains
\begin{align}
b_1=-\frac{2\nk^2+\nq_z(4+3\nq_z)}{4(\nq_z+1)}~,
%\label{eq:}
%
\end{align}
so the first pole-skipping point is given by $(\nq_z, \nk^2)=(-1,1/2)$. Namely, one cannot see the hydrodynamic pole-skipping $(\nq_z, \nk^2)=(0,0)$.

Similarly, for the gravitational vector mode, there are 2 gauge-invariant variables $\mfh_{zy}, \mfh_{uy}$, and we choose $\mfh_{zy}$. If one chooses $\mfh_{uy}$ as the master variable, the master equation is given by
\begin{align}
0= Z_2''+\left( \frac{f'}{f}+\frac{2}{u} \right) Z_2'-\frac{\nq_z^2+\nk^2f}{uf^2}Z_2~,
%\label{eq:}
%
\end{align}
where $Z_2=f\mfh_{uy}$.
%Asymptotically, $Z\sim u^\lambda$ with $\lambda=-1,0$ ($\lambda$ should differ by 2?).
In this case, one obtains
\begin{align}
b_1=-\frac{-4+2\nk^2+2\nq_z+3\nq_z^2}{4(\nq_z+1)}~,
%\label{eq:}
%
\end{align}
so the first pole-skipping point is given by $(\nq_z, \nk^2)=(-1,3/2)$, and again one cannot see the hydrodynamic pole-skipping.

\end{document}